\documentclass[11pt,a4paper]{article}
\usepackage[utf8]{inputenc}
\usepackage[T1]{fontenc}
\usepackage{amsmath,natbib}
\usepackage{amssymb}
\usepackage{makeidx}
\usepackage{graphicx,color}
\usepackage{comment}

\begin{document}
\title{Dependence  of Solar supergranular lifetime on surface magnetic activity and rotation}
\author
{Sowmya G M$^1$,Rajani G$^2$, U Paniveni$^3$, R Srikanth$^3$}
\date{}
\maketitle
$^1$ GSSS Institute of Engineering and Technology for Women, KRS Road, Metagalli Mysuru-570016, Karnataka, India\\
$^2$ PES College of Engineering, Mandya - 571401, Karnataka, India.\\
$^3$ Poornaprajna Institute of Scientific Research,Devanahalli, Bangalore-562110, Karnataka, India\\

\begin{abstract}
The lifetimes and length-scales for supergranular cells in active and quiescent regions of the Solar chromosphere, and the relation between the two, were studied using a time series of Ca II K filtergrams.  The lifetimes, in contrast to supergranular length scale and fractal dimension, show no significant dependence on Solar latitude, suggesting that cell lifetimes are independent of the differential rotation and a possible supergranular super-rotation. The functional form of the relation was obtained guided by a comparison of the distributions of the two supergranular parameters.  We infer a linear dependence of cell lifetime on area, which can be understood by the assumption of the network's evolution via a diffusion of the magnetic field. Our analysis suggests that the diffusion rate in quiet regions is about 10\% greater than in active regions.
\end{abstract}


\section{Introduction}
The supergranular network is the superficial manifestation of Solar convection and is important for solar flux transport. The existence of a strong correlation between the chromospheric networks and the supergranulation structure was pointed out first by Leighton on the basis of Dopplergrams \citep{leighton1962velocity,leighton1963solar}.  Subsequent studies made use of Ca II K spectroheliograms and then filtergrams, an important tool to probe Solar convection and also magnetism \citep{chatzistergos2022full}. Supergranular network cells (called ``supergranules'') are characterized by distributions centered around a lifetime $T \approx 25$ hours and length scale $L \approx 35$ Mm. Since the early work of \citet{simon1964velocity}, these two parameters along with the velocity and magnetic fields associated with supergranules have been studied and reported in a wide range of values by various workers, cf. \citet{mcintosh2011observing, mandal2017association, chatterjee2017variation, rajani2022solar} and reference therein.  More recently, space borne instruments such as Helioseismic Magnetic Imager (HMI) on board Solar Dynamics Observatory (SDO) and Solar and Michelson Doppler Imager (MDI) on board Heliospheric Observatory (SOHO) \cite{williams2014analysis} have been used to study supergranulation.

The derived lifetimes of supergranulation shows good dependence on the choice of method and region.
\citet{janssens1970long}, using H$\alpha$ filtergrams, estimated that $T\approx 21$ hours. \citet{livingston1974magnetic} obtained a similar value of 22 hour and this was also reported by \citet{singh1994study} based on the observation of the appearance and disappearance of the cell. On the other hand, the same authors observed that certain exceptional supergranules found in the vicinity of active regions can survive for several days. By visual examination of individual supergranules \citet{wang1988velocity} detected $T > 50$ hour, and pointed out that this can be larger than that obtained by the cross-correlation (CC) method \citep{rogers1970lifetime}, owing to the latter's sensitivity to  shape changes. \citet{simon1964velocity} estimated a CC lifetime of 20 hours using a time series of Ca K spectroheliograms. \citet{worden1976study} employed the CC technique to estimate a lifetime of 36 hours using magnetogram data. Observing the Fe I 8988\AA  ~and Ca II K networks,  \citet{duvall1980equatorial} and  \citet{raju1998dependence} derived a CC lifetime of 42 hr and 25 hr, respectively.   

Visual inspection techniques possess the advantage of being able to directly follow intricate morphological changes such as merging, splitting, migration, disappearance, and appearance of magnetic fluxes that make up the evolution of the network \citep{harvey1973ephemeral, wang1995flux}. Therefore, lifetime estimated using this would better reflect the processes underlying supergranular dynamics. By contrast, correlation techniques fail to distinguish between true aspects of cell evolution such as the disappearance or appearance of certain features and shape changes arising from the relocation of  magnetic elements. 

On the question of whether a supergranule survives beyond its correlation lifetime, there has been conflicting evidence:  \citep{wang1988velocity} reports in the affirmative, but comparable results found by \citet{rogers1970lifetime} and \citet{janssens1970long}, using the correlation and morphological techniques for H$\alpha$ data,  respectively,  report in the negative. Similarly, lifetimes estimated by \citet{raju1998correlation} and \citet{singh1994study,paniveni2010activity} on Ca II K using correlation and visual inspection techniques show comparable results for both active and quiescient network regions.  In the case of extended supergranular networks, correlation lifetimes can frequently assume values  as large as 45-60 hours. However, in contexts involving intricate morphological changes, as for example the long-lived features such as magnetic \textit{pukas} or plages \citep{livingston1974magnetic}, it is more advantageous to study lifetimes visually. 

In this work, we investigate and contrast the relations between lifetimes and length-scales for supergranular cells in active and quiescent regions of the Solar chromosphere was studied using a time series of Ca II K filtergrams, extending work done by \citet{singh1994study} and \citet{srikanth1999chromospheric}.  Here it may be noted that the Ca II K network traces out magnetic flux concentrations at the supergranular boundary thanks to the enhanced network brightness they produce \citep{spruit1990solar,hagenaar1997distribution,raju2002dependence}. Based on a data analytic method proposed by the latter, we deduce the functional form with guidance by the distributions of the two supergranular parameters.  Our results are found to support the expected picture of the network's evolution through a diffusion of the magnetic fields, and the influence of the fields on cell properties.

The paper is structured as follows. In Section \ref{sec:data}, we introduce the data used and the method of its analysis. The basic results for cell lifetime and length scale are presented in Section \ref{sec:results}.

\section{Data and Analysis \label{sec:data}}
\begin{figure}
	\centering
\includegraphics[width=3in]{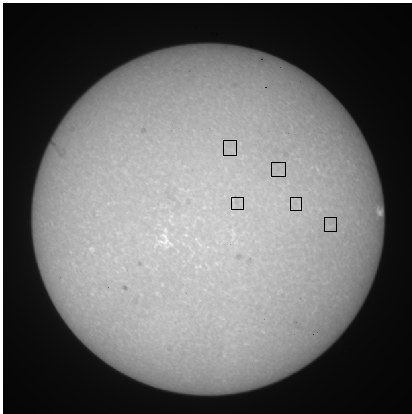}
\caption{Quiescent region cells: Selection of supergranules in quiet regions of the Solar chromosphere; from the KSO archive of the 23rd cycle.}
\label{fig:quiet}
\end{figure}

\begin{figure}
	\centering
	\includegraphics[width=3in]{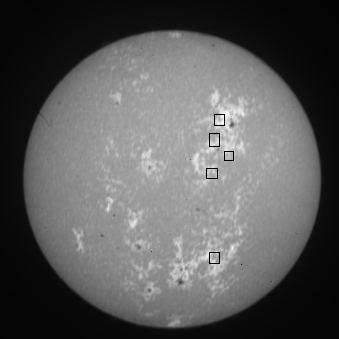}
	\caption{Active region cells: Selection of supergranules in active regions of the Solar chromosphere; from the KSO archive of the 23rd cycle.}
	\label{fig:active}
\end{figure}

\begin{figure}
	\centering
	\includegraphics[width=3in]{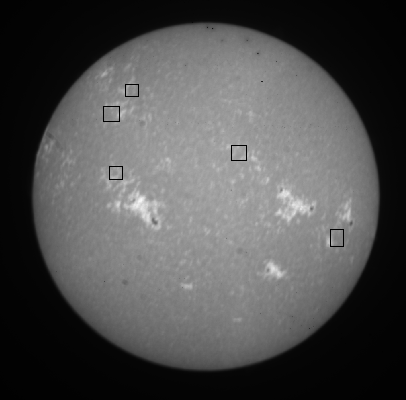}
	\caption{Semi-active region cells: Selection of supergranules in semi-active regions of the Solar chromosphere; from the KSO archive of the 23rd cycle.}
	\label{fig:semi}
\end{figure}

Supergranular size $L$ was estimated as square root of  the area enclosed within the cell boundaries traced out on the Ca II K filtergrams. For lifetime estimation, observations were made for the Kodaikanal Solar Observatory (KSO) data for the year 1998, 2002, 2004 and 2007 for the descending, minimum and active phases. In this analysis used data consisting of approximately 1200 Ca II K filtergrams of the 23rd Solar cycle. Time averaging over 10 min is used in order to eliminate noise due to 5-min oscillations. This method yields approximately six data-frames per hour.  To estimate lifetime at a given epoch, approximately 72 hours of data are considered, which span about 432 frames at 10-minute inter-frame intervals. A specific  supergranular cell is tracked across frames sequentially, with lifetime being estimated as the time interval between the frame of its initial appearance and that of its last disappearance \citep{paniveni2004relationship}. Supergranular lifetime has been estimated for quiescent, active and semi-active regions. By quiet region cells, we mean those found far from the magnetically active regions (see Figure \ref{fig:quiet}). Active region cells are found in close proximity of active regions (see Figure \ref{fig:active}), whilst the semi-active region supergranules are found in regions of intermediate magnetic activity (see Figure \ref{fig:semi}). 

In previous studies supergranular lifetime was obtained via cross-correlation applied to time series data \citep{srikanth1999chromospheric}. The behaviour was analysed assuming the diffusion elements of the magnetic network with observed lifetime identified an a diffusion time-scale. By contrast, lifetime estimates based on visual inspection, as done here, are implicitly related to the crossing time of the plasma from the cell center to its edge \citep{krishan1999two}. Therefore, visual inspection is expected to yield the eddy turnover time. While lifetime estimate by visual inspection is rather tedious, still it is fairly reliable \citep{paniveni2010activity}. Our sample size is small, but brings out characteristic features contrastng the different activity epochs and regions.

Our work here has been focused on studying the behaviour of supergranules in quiet, intermediate and active regions.  The contrast in the cell lifetime across regions of different activity levels may be theoretically modelled as arising out of differences in diffusion rates of the magnetic flux transport \citep{schrijver1989relations}.
 \section{Results \label{sec:results}}
   
\subsection{Cell Lifetime dependence on Solar latitude}
   
Our estimates of cell  lifetime in quiet region cells are comparable to those previously reported based on the KSO data \citep{chatterjee2017variation, mandal2017association, sowmya2022supergranular, rajani2022solar}. For example, the estimate for quiet and semi-active regions in this data match that obtained by \citet{singh1994study}, who find $T \in [15,40]$ hour, with the most likely lifetime being 22 hours. The active region lifetime is estimated by those authors to be almost double the quiescent value, in agreement with our result as indicated in Table \ref{tab:lifetime}. 
   
\begin{figure}
   \centering
   \includegraphics[scale=0.75]{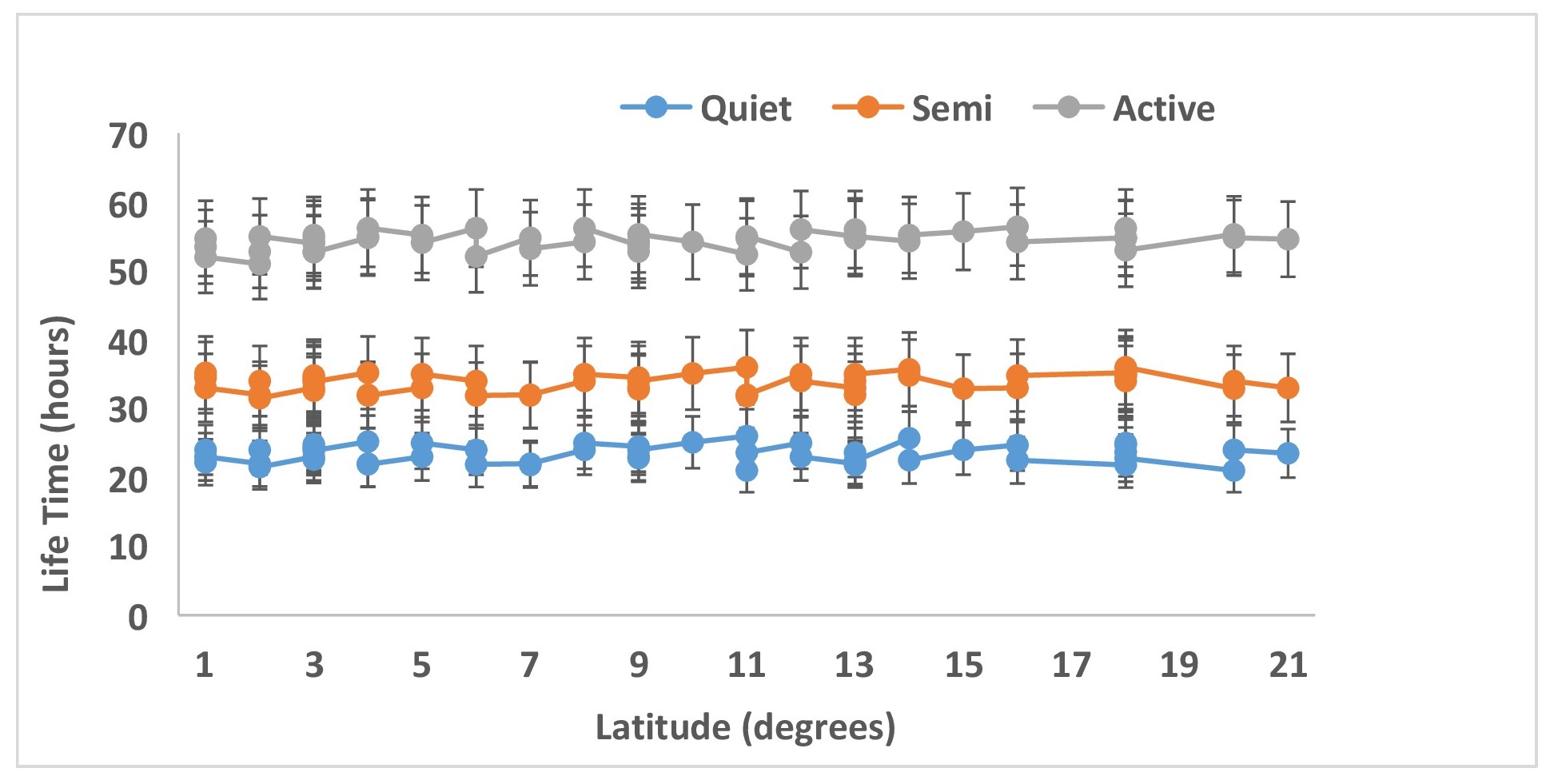}
   \caption{Plot of supergranular lifetime (in hours) with respect to Solar latitude (in degrees) using data of the 23rd Solar cycle.}
   \label{fig.1} 
   \end{figure}

A plot of cell lifetimes versus latitude for the data points is shown in Figure~\ref{fig.1}.   This stands in contrast to length scale and fractal dimension. The latter shows a latitude dependence, which may be potentially linked to the Sun's differential rotation \cite{sowmya2022supergranular}.  Supergranular scale shows  the vertical-horizontal asymmetry at higher latitudes \citep{raju2020asymmetry}. Thus, our observation here suggests that cell lifetime is unaffected by Solar rotation, whereas spatial properties of cells indeed manifest an influence.

We may consequently also rule out any dependence of supergranulation lifetime on superrotation, the possible faster rotation of supergranules with respect to magnetic structures and plasma. However, it may be noted that superrotation may well be an artefact of projection effects in Dopplergrams \citep{meunier2007superrotation}.
 
 \begin{table}
	\centering
	\begin{tabular}{| c | c |}
		\hline
		Region  &    Lifetime(hour)\\
		\hline
		Quiet   &   23.58 $\pm$ 1.3\\
		\hline
		Semi Active   &   34 $\pm$ 1.7\\
		\hline
		Active   &   54.4 $\pm$ 1.6\\
		\hline
	\end{tabular}
	\caption{Solar supergranular lifetime over the quiet, semi and active regions.}
	\label{tab:lifetime}
\end{table}


 \subsection{ Deducing the functional dependence of cell lifetime on the size \label{sec:skew}}
\begin{figure}
 	\centering
 	\includegraphics[scale=0.75]{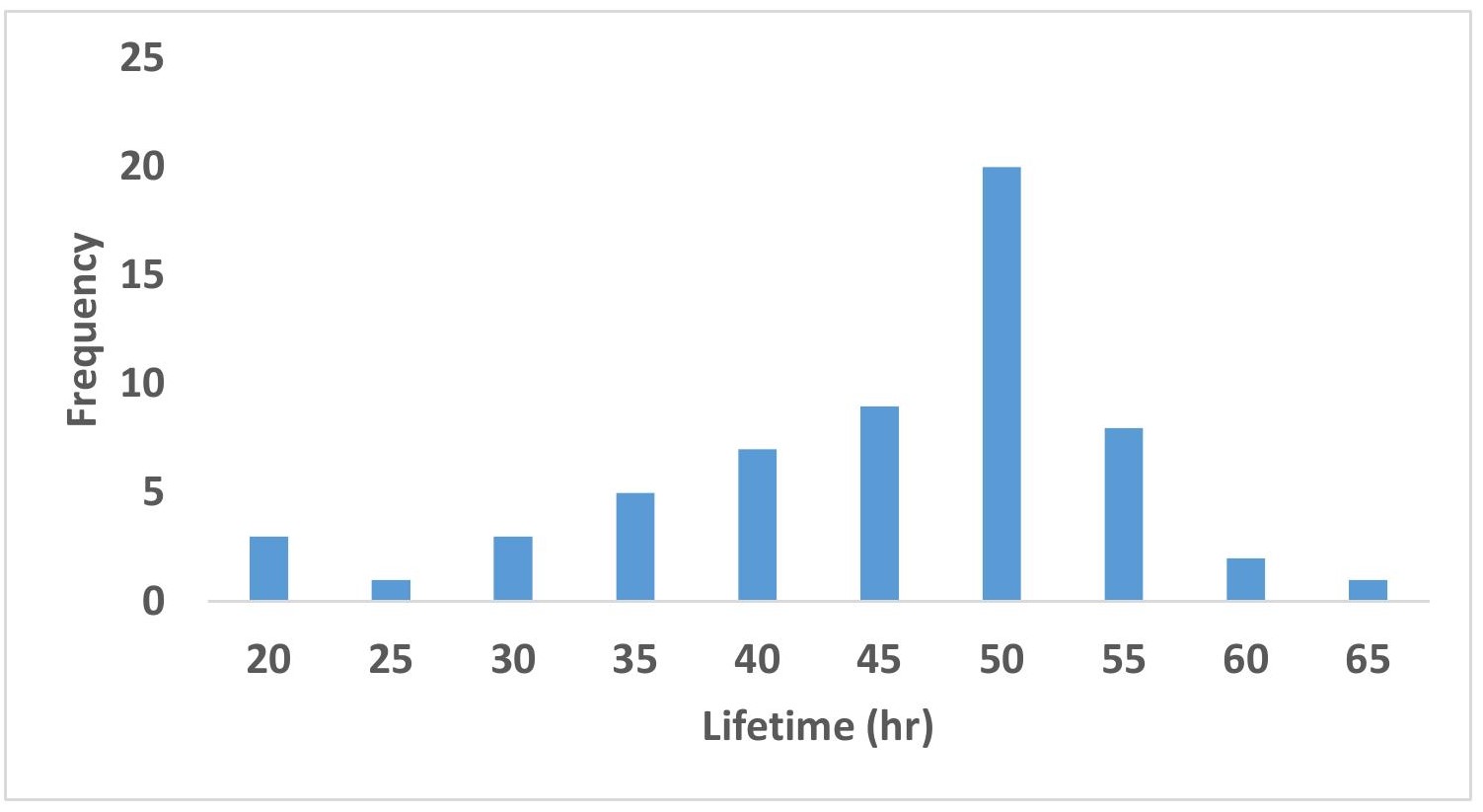}
 	\caption{Histogram of lifetime of the Ca II K network cells in the active region. The curve shows a left-hand side tail.
 		The skewness and kurtosis derived for this distribution are given in Table \ref{tab:LS}.}
 	\label{fig:histoLA}
 \end{figure}

\begin{figure}
 	\includegraphics[scale=0.75]{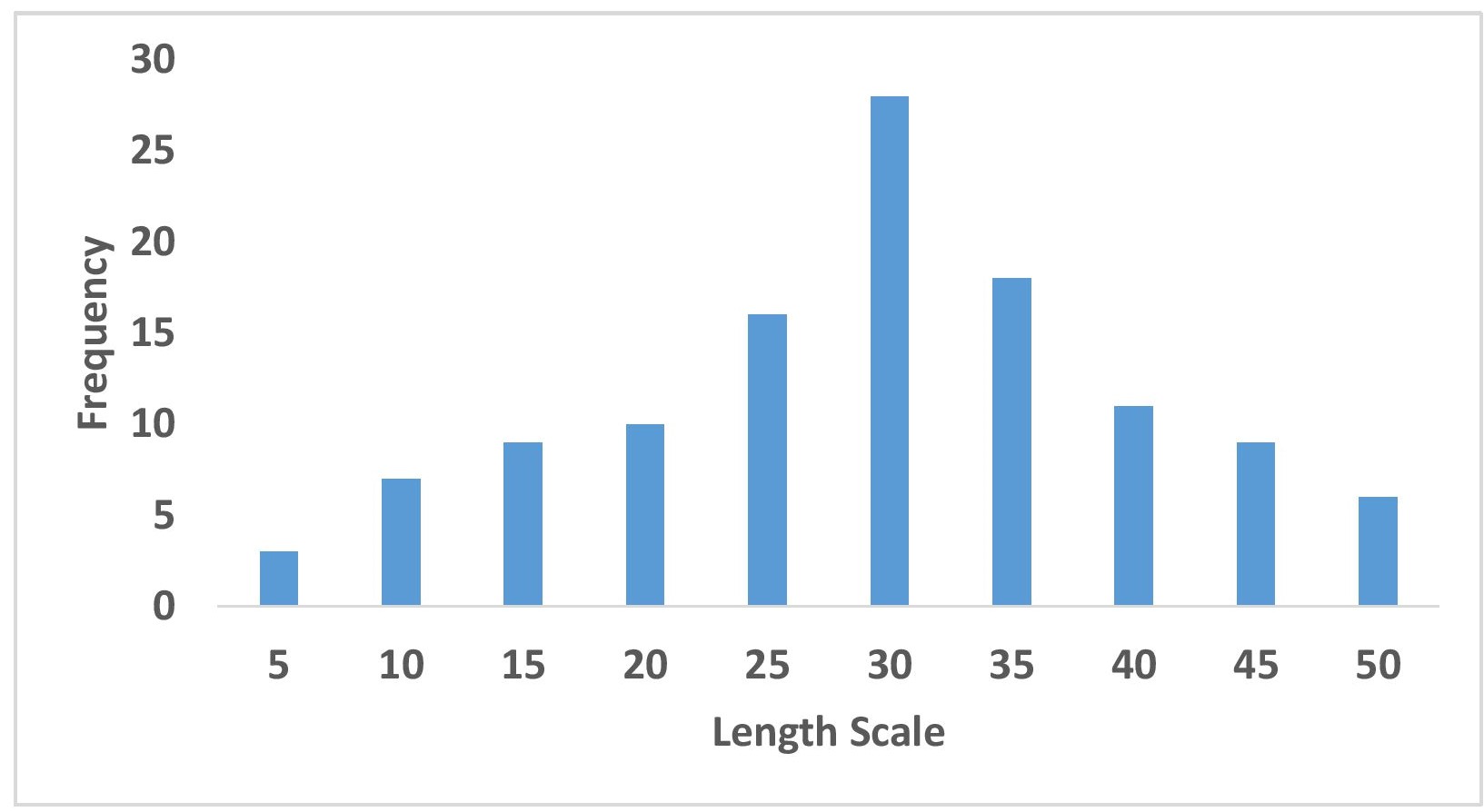} 
 	\caption{Histogram of length scale of supergranules in the active region. The skewness and kurtosis derived for this distribution are given in Table \ref{tab:LS}.}
 	\label{fig:histoSA}
 \end{figure}

 The parameters such as cell lifetimes $T$ and Length scales $L$, are evidently interdependent. One can directly estimate the functional dependence of $L$ on $T$ by  curve fit algorithm. We also expect this dependence to be reflected in the distribution of these two parameters, given in Figures \ref{fig:histoLA} and \ref{fig:histoSA}. This information can also be used to help with the  estimation of the functional relation between $L$ and $T$ for the quiet or active region.  
		 
Here we employ an indirect method to estimate the functional relationship between $L$ and $T$ by estimating the transformation that would map the distribution of the latter with respect to that of the former. Somewhat simplistically, we shall assume that their respective distribution is sufficiently represented by two parameters: (1) the skewness $\varsigma$, which is a measure of asymmetry of the distribution about the mean $\mu$, and (2) the kurtosis $\kappa$, which is a measure of the ``tailedness'' in the distribution. 

In statistics, given a random variable described by probability distribution $f(x)$, skewness is a quantification of asymmetry of $f(x)$. It is given by the third standardized moment, defined as follows: 
\begin{equation}
\begin{aligned}
\centering
\varsigma=\frac{1}{\alpha^3}\int_{-\infty}^{\infty}(x -\mu)^3 f(x) dx,
\label{eq:skew}
\end{aligned}
\end{equation} 
where $\alpha$ is the standard deviation. A distribution may be right-skewed (resp., left-skewed), when it has a more prominent tail on the positive (resp., negative) side about the mean. A zero-skew distribution is perfectly symmetric on both wings about the mean. As basic examples, a normal distribution has zero skewness; an exponential distribution has skewness $\varsigma =2$, and for a lognormal distribution describing a random variable $X$ whose logarithm $\ln(X)$ is described the normal distribution with variance $\beta$, we have $\varsigma = (e^{\beta}+2)\sqrt{e^{\beta}-1}$.

Given a random variable described by probability distribution $f(x)$, kurtosis is a quantification of how tailed $f(x)$ is, i.e., how well the distribution features outliers in the extreme, rather than concentration of data closer to the mean. It is given by the fourth standardized moment, defined as follows
\begin{equation}
\begin{aligned}
\centering
\kappa=\frac{1}{\alpha^4}\int_{-\infty}^{\infty}(x -\mu)^4 f(x) dx.
\label{eq:kurt}
\end{aligned}
\end{equation}
As basic examples, a normal distribution has kurtosis $\kappa=3$; a Laplace distribution has $\kappa =6$, and the uniform distribution has  $\kappa=1.8$. A distribution may be platykurtic (resp., leptokurtic), when it has lesser (resp., greater) kurtosis than the normal distribution. The skewness and kurtosis for our data is summarized in Table \ref{tab:LS}.
   
\begin{table}
   \begin{tabular}{| c | c | c |}
   \hline
   Field  & For Length scale distribution  & For Lifetime distribution\\
   \hline
   Skewness  &  (0.463, 0.779)   &  (0.865, 1.278)\\
   \hline
   Kurtosis  &  (2.75, 2.638)  &  (2.045, 2.028)\\
   \hline
   \end{tabular}
   \caption{Statistics of lifetime and scale distribution for the Ca II K networks cells for (active, quiet) regions. We note that all distributions are platykurtic, i.e., having a lower kurtosis than a normal distribution for which kurtosis $\kappa=3$.}
   \label{tab:LS}
   \end{table}

When a random variable $X$ is subjected to a transformation $\eta$, then the properties of the distribution of the transformed variable $\eta(X)$, in particular $\varsigma$ and $\kappa$, will in general be different from those of the distribution of $X$. For example, above we saw that whereas the normal distribution of some random variable $X$ has zero skewness, the lognormal distribution (which characterizes $e^X $) is skewed positively. This means that if we let cell lifetimes and scale to be related by $T \equiv \eta(L)$, then the right $\eta$ will ensure that the skewness and kurtosis of $\eta(L)$ is close to the corresponding values of the distribution of $T$. Obviously infinitely many functions $\eta$ may satisfy this requirement. We must thus restrict $\eta$ to a reasonable family of two parameters for this approach to work. This is done as follows.
  
Under an invertible transformation $\eta$ of the random variable $x$ given by $y \equiv \eta(x)$, let the distribution function $f(x)$ become $g(y)$, which is determined as follows. By  definition:
\begin{equation}
\begin{aligned}
\centering
 \int_{x_{1}}^{x_{2}} f(x) dx = \int_{\phi(x_1)}^{\phi(x_2)} g(y) dy,
  \label{eq:integral}
 \end{aligned}
 \end{equation}
owing to conservation of probability.
The positive definiteness of $f(x)$ implies that:
$
  g(y)\vert dy\vert = f(x)\vert dx\vert,
$
whereby
   \begin{equation}
   \begin{aligned}
   \centering
   g(y) = f[\eta^{-1}(y)] \vert (\eta^{-1})^{\prime}(y)\vert,
   \label{eq:eq5}
   \end{aligned}
   \end{equation}
where the prime symbol denotes the first derivative.
   
Let $T = \eta(L)$. Further, let $f(L)$ and $g(T)$ denote the respective distribution function. In order for our method to work, we must restrict to a two-parameter family of transformations. It is reasonable to confine $\eta$ to a polynomial relation of the form
\begin{equation}
   \begin{aligned}
   \centering
  		 T = aL^ n + b 
  		\label{eq:function}
   \end{aligned}
   \end{equation} 
  		 According to equation (\ref{eq:eq5})
 \begin{equation}
   \begin{aligned}
   \centering 		 
  		g(T) = \frac{f(L)}{\left[(L-b)^{n-1}a\right]^\frac{1}{n}n} 
 \label{eq:eq7}
   \end{aligned}
   \end{equation}
We now apply this exercise to our lifetime vs length scale data.
 
Table ~\ref{tab:LS} summarizes the skewness and kurtosis data for the active region given in histograms Figures~\ref{fig:histoLA} and ~\ref{fig:histoSA}, and additionally for quiet regions (not included, for brevity). Based on the skewness, we find that in either region, supergranular scales are less asymmetric than lifetimes. This feature seems generic for supergranules, irrespective of the activity level. 
 
\begin{figure}
	\includegraphics[scale=0.75]{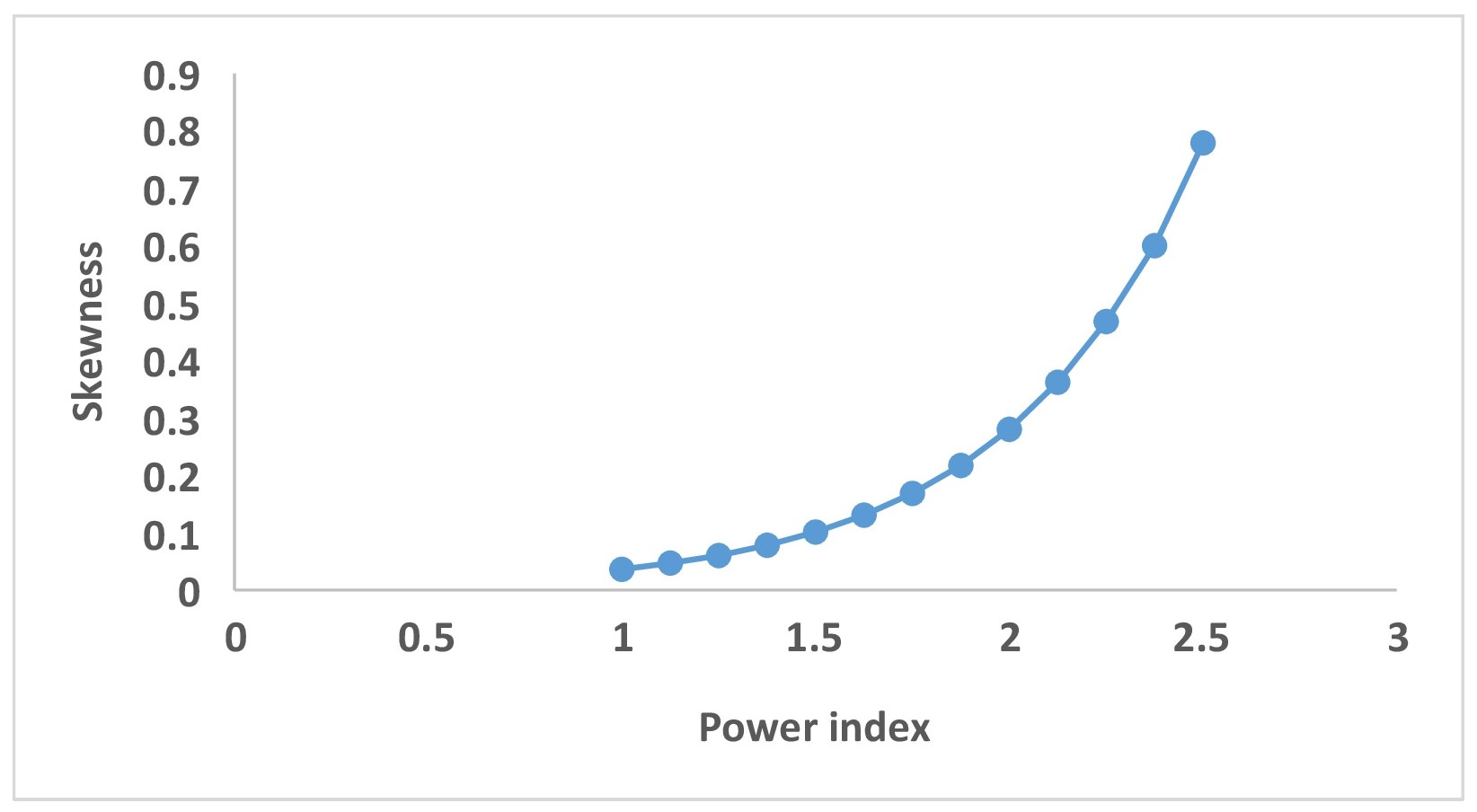}
	\caption{Plot of skewness of distribution for various powers indices of length scale. The skewness of the 
		lifetime distribution is 0.865.}
	\label{fig:skew}
\end{figure}

\begin{figure}
	\includegraphics[scale=0.75]{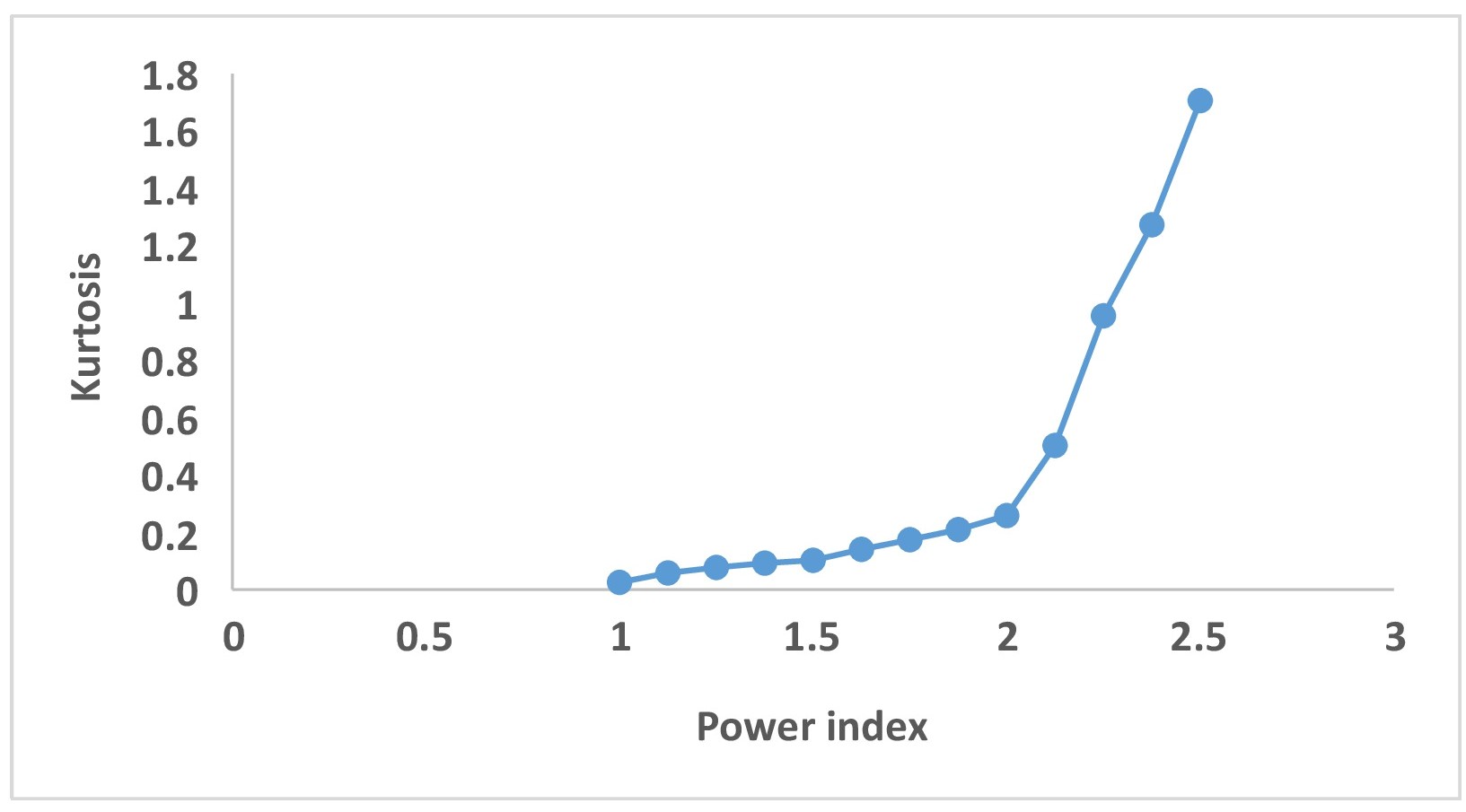}
	\caption{Plot of kurtosis distribution for various powers 'n' of length scale}.
	\label{fig:kurtosis}
\end{figure}

To determine $n$ for a given region (active or quiet), the values of skewness and kurtosis for the transformed length scale are plotted as a function of $n$ in the range $1.0 \leq n \leq 2.5$. The plots for the two, namely  $\varsigma(n)$ and $\kappa(n)$, are given in Figures~\ref{fig:skew} and ~\ref{fig:kurtosis}, respectively. As expected, both plots exhibit a monotonic increase for $x \geq 1.0$. 

\begin{figure}
	\centering
	\includegraphics[scale=0.75]{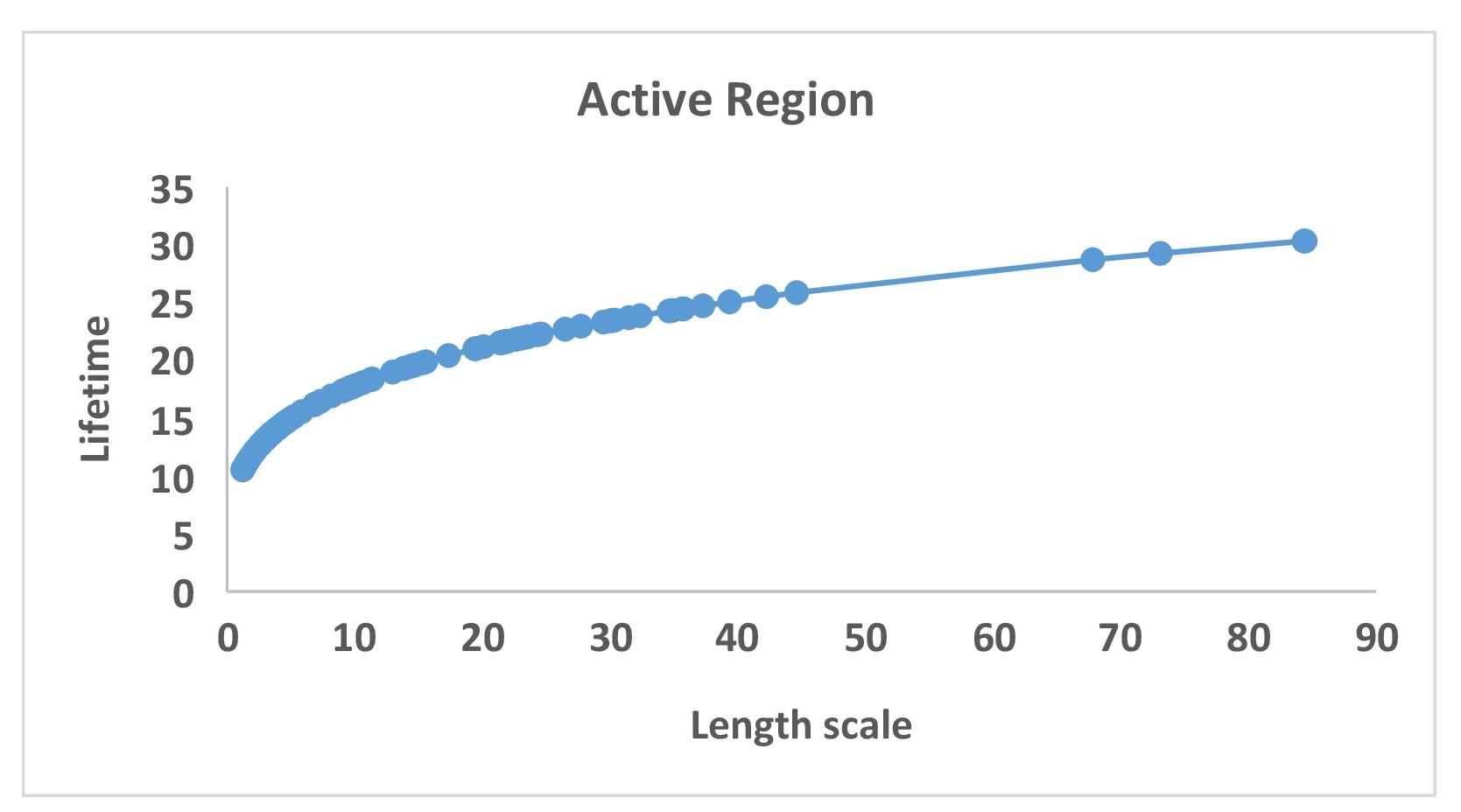} 
	\caption{Variation of lifetime as a function of length scale for Ca II K network cells for active region. The fit function is given by Eq. (\ref{eq:activeTL})}
	\label{fig:activefit}
\end{figure}

In the case of active regions, for the observed lifetime distribution skewness $\varsigma$ = 0.463 (Table \ref{tab:LS}), the skewness for length scale distribution corresponds to the Figure~\ref{fig:skew}, range 2.125$\leq$ n $\leq$ 2.25. Similarly, for the observed lifetime distribution, kurtosis $\kappa =2.75$, the kurtosis of the transformed length scale distribution corresponds to the range $1.875 \leq n \leq 2$ as shown in Figure~\ref{fig:7} . We choose $n=2$, and the Monte-Carlo least-squares curve fitting algorithm yields the function:
\begin{equation}
T = 7.45 + 3.5  A,
\label{eq:activeTL}
\end{equation}
where $T$ is given in hours and supergranular area $A \equiv L^2$ is in units of Mm$^2$. More specfically, indicating error bars, we may give in place of Eq. (\ref{eq:activeTL}), the fit function $T = \alpha + \beta A$, where $\alpha$ and $\beta$ are the fit constants having the units of $T$ and $TL^{-2}$, respectively, with $\alpha = 7.45 \pm 0.025$ hours and $\beta=3.5 \pm 0.01$ Mm$^2$. Figure \ref{fig:activefit} depicts the observed data on the dependence of lifetime on scale in active regions, as well as the fit based on Eq. (\ref{eq:activeTL}). This shows a reasonably good agreement between the two.

\begin{figure}
	\centering
	\includegraphics[scale=0.75]{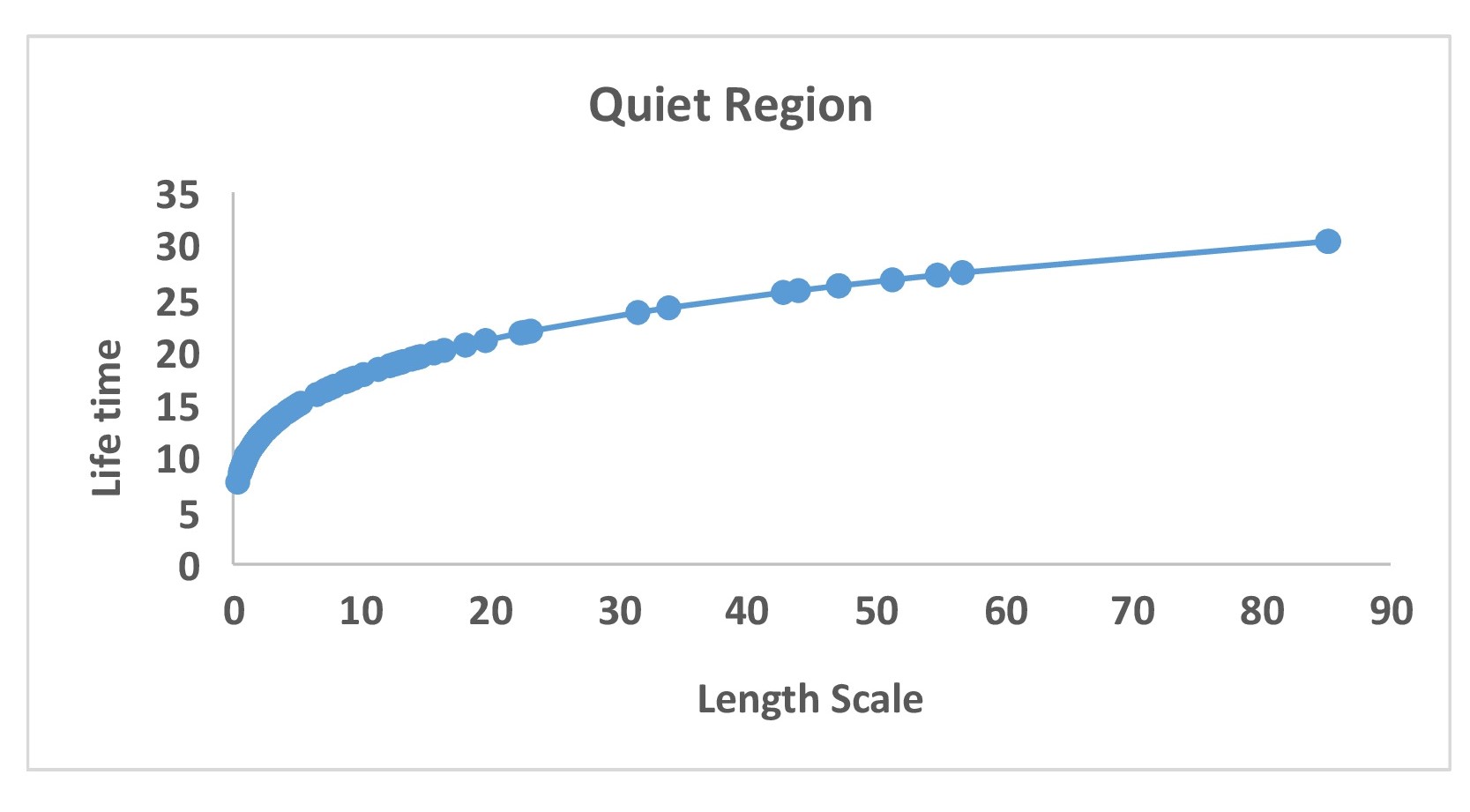} 
	\caption{Variation of lifetime (hr) as a function of length scale for Ca II K network cells for quiet region. The plot is a fit using the functional form of Eq. (\ref{eq:function}).}
	\label{fig:quietfit}
\end{figure}

For quiet regions, the analogous calculation yields, in case of the skewness data of of length scale and lifetime distributions, the range $2 \le n \le 2.125$., and in the case of kurtosis of the length scale and lifetime distributions, the range
$1.75 \le n \le 2$. Analogous to Eq. (\ref{eq:activeTL}), in the quiet region, the Monte-Carlo curve fitting algorithm yields the function:
\begin{equation}
T = 6.75 + 3.25  A,
\label{eq:quietTL}
\end{equation}
where $T$ is given in hours and $L^2$ in units of Mm$^2$. As before, indicating error bars, we may give in place of Eq. (\ref{eq:quietTL}), the fit function $T = \mu + \nu A$, where $\mu$ and $\nu$ are the fit constants having the units of $T$ and $TL^{-2}$, respectively, with $\mu = 6.75 \pm 0.023$ hours and $\nu=3.25 \pm 0.02$ Mm$^2$.
  Figure \ref{fig:quietfit} depicts the observed data on the dependence of lifetime on scale in active regions, as well as the fit based on Eq. (\ref{eq:quietTL}). This shows a reasonably good agreement between the two.

Eqs. (\ref{eq:activeTL}) and (\ref{eq:quietTL}) are consistent with the dependence reported by \citet{singh1994study, srikanth1999chromospheric}, where the authors find a linear relation between lifetime and scale of supergranules, which can broadly be understood through a model where cell lifetime is related to the diffusion of magnetic elements.
 However, the above authors restrict their study only to quiet regions, whereas we extend the study to a comparative study of quiet and active regions. 

\section{Discussion and Conclusion}
We studied the lifetimes and length-scales of supergranular cells in active and quiescent regions of the Solar chromosphere, and the relation between the two,  using a time series of Ca II K filtergrams. We find that the lifetimes show no significant dependence on Solar latitude, suggesting that cell lifetimes are independent of the differential rotation. This independence stands in contrast to supergranular length scale and fractal dimension. For example, \citet{raju1998dependence} have noted that supergranular size as observed in the Ca II K  shows upto a 7\% latitudinal variation. \citet{sowmya2022supergranular} report an anticorrelation between fractal dimension and latitude in the belt between 20$^{\circ}$ N and 20$^{\circ}$ S.

Our results on the lifetime-scale relation can be interpreted to shed light on the relative dynamics of the active and quiet regions of the Sun. From Eqs. (\ref{eq:activeTL}) and (\ref{eq:quietTL}), we find that the slope  is 3.5 hr  Mm$^{-2}$, which is slightly larger in the case of active regions than in the quiet regions, namely 3.25 hr  Mm$^{-2}$. This difference may be understood as a consequence of the fact that lifetime and scale of an active or quiet region cell can depend on its interaction with the ambient magnetic fields.  In specific, the above noted difference in slopes may be attributed to two related factors: (a) lowering of cell size in the presence of magnetic activity \citep{singh1982}, and (b) the enhancement of cell lifetime in active regions, as noted in Table \ref{tab:lifetime}. The effect of magnetic flux  can be understood as due to plasma confinement by the magnetic field \citep{sowmya2022supergranular}. 

Accordingly, the slope $dT/dA$ in  Eqs. (\ref{eq:activeTL}) or (\ref{eq:quietTL}) may be interpreted as the inverse of the diffusion coefficient $D$  associated with the cell, i.e., as $1/D$. For active regions, we then have $D = 10^6/(3.5  ~[\pm 0.01] \times 3600) \approx 79.3 \pm 0.2 $ km$^2$/s. Similarly, we obtain about $D \approx 85.5 \pm 0.5$ km$^2$/s for quiet regions. Intuitively, the longer lifetime of cells in the active region is due to the lower diffusion rate, and our results imply that in the active region, the diffusion happens about $79.3/85.5 \approx 0.93$ slower than in the quiet region. It may be noted that this is in agreement with recent works \citep{abramenko2017dispersion, abramenko2018dispersion} demonstrating superdiffusivity in quiet regions and nearly-normal diffusion in active regions. Specifically, the pattern of greater diffusivity in quiet regions than in active ones is found to be pronounced in the case of cells with scale greater than 5 Mm \citep{abramenko2018dispersion}, which is compatible with the scale range appropriate to the present data.

As such the relative diffusion of the active and quiet regions is expected to be dependent on the phase of the cycle. For example, during solar minimum a large cell in the active region may retain its identity for upto 10 months, where during solar maximum, the exceptional large lifetimes rarely exceed 4 months. These issues may be studied in the future in contnuation of work reported here.

   \bibliographystyle{apalike}
   \bibliography{./References}
    \end{document}